\documentclass[aps,prl,twocolumn,nofootinbib,superscriptaddress,tightenlines]{revtex4}

\usepackage{amsmath,amsthm,latexsym,amssymb,amsfonts}
\usepackage{graphicx}
\usepackage{hyperref}

\def\be{\begin{equation}}
\def\ee{\end{equation}}
\def\ba{\begin{eqnarray}}
\def\ea{\end{eqnarray}}
\def\A{\text{A}}
\def\B{\text{B}}
\def\C{\text{C}}
\def\D{\text{D}}
\def\SU{\text{SU}}
\def\bX{\mathbf{X}}
\newcommand{\g}[1]{g^{\vphantom{-1}}_{\vphantom{1'}\scriptstyle #1}}
\newcommand{\gm}[1]{g^{-1}_{\vphantom{1'}\scriptstyle #1}}
\newcommand{\h}[1]{h^{\vphantom{-1}}_{\vphantom{1'}\scriptstyle #1}}
\newcommand{\hm}[1]{h^{-1}_{\vphantom{1'}\scriptstyle #1}}
\newcommand{\bd}{\mathbf d}
\newcommand\nn{\nonumber}

\begin{document}

\title{A new vacuum for Loop Quantum Gravity}

\author{Bianca Dittrich}
\affiliation{Perimeter Institute for Theoretical Physics,\\ 31 Caroline Street North, Waterloo, Ontario, Canada N2L 2Y5}

\author{Marc Geiller}
\affiliation{Institute for Gravitation and the Cosmos \& Physics Department,\\ Penn State, University Park, PA 16802, U.S.A.}

\begin{abstract}
We construct a new vacuum for loop quantum gravity, which is dual to the Ashtekar--Lewandowski vacuum. Because it is based on BF theory, this new vacuum is physical for $(2+1)$--dimensional gravity, and much closer to the spirit of spin foam quantization in general. To construct this new vacuum and the associated representation of quantum observables, we introduce a modified holonomy--flux algebra which is cylindrically consistent with respect to the notion of refinement by time evolution suggested in \cite{refining}. This supports the proposal for a construction of a physical vacuum made in \cite{bd12b,refining}, also for $(3+1)$--dimensional gravity. We expect that the vacuum introduced here will facilitate the extraction of large scale physics and cosmological predictions from loop quantum gravity.
\end{abstract}

\maketitle

\section*{1. Introduction}

Loop quantum gravity \cite{books} is a background independent approach to the quantization of gravity. As such, it is based on a (kinematical) vacuum, the Ashtekar--Lewandowski vacuum \cite{al}, which can be seen as fundamental since it is peaked on the totally degenerate spatial geometry and has maximal uncertainty in the conjugated variable. This makes however the extraction of large scale physics quite difficult, as one has first to build up a ``condensate'' state that describes a smooth geometry at larger scales. Alternative vacua, which include non--degenerate geometries and have similar peakedness properties to the Ashtekar--Lewandowski vacuum, have been defined in \cite{koslowski}. Due to the introduction of a specific spatial metric on which the non--degenerate geometries are peaked, these states are not diffeomorphism--invariant anymore (see however \cite{KS-vacuum} for attempts to define diffeomorphism invariance). Here we sketch the construction of a vacuum state that is dual to the Ashtekar--Lewandowski vacuum and carries a notion of spatial and space--time diffeomorphism invariance. Since this vacuum is based on BF theory, which underlies the spin foam approach, we hope that it will more directly connect loop quantum gravity to spin foam quantization \cite{SFreviews}, and moreover facilitate the derivation of large scale physics.

The Ashtekar--Lewandowski (AL) vacuum is central for the definition of the continuum Hilbert space of loop quantum gravity (LQG hereafter), which arises from a refinement (inductive) limit of a family of Hilbert spaces based on graphs. The AL vacuum is ($a$) a diffeomorphism--invariant state which is peaked on a spatial geometry that is totally degenerate and has maximal uncertainty in the conjugated holonomy variables. The vacuum also determines the  ($b$)  embedding of states based on a given graph into a Hilbert space based on arbitrarily refined graphs. Indeed, the additional degrees of freedom on the refined graph are put into the vacuum state. Furthermore ($c$) the vacuum is cyclic, which means that the LQG Hilbert space is spanned by states generated from the vacuum by (cylindrically consistent) observables, in this case holonomy observables. 

This vacuum is often called kinematical since one expects that physical states satisfying the Hamiltonian and diffeomorphism constraints will be based on a physical vacuum describing rather a non--degenerate vacuum. The quest for other vacuum states \cite{sahlmann1} has led to the F--LOST \cite{FLOST} uniqueness theorem for LQG, that states that the AL vacuum is unique given a certain number of assumptions. In particular, one requires a representation of the so--called holonomy--flux algebra of LQG together with the diffeomorphism invariance of the vacuum.

Alternatively, one might ask for a vacuum which already incorporates the dynamics of the system. Indeed, the vacuum proposed in the present work is motivated by considerations in \cite{bd12b,refining}, where a construction principle for the physical vacuum is provided. As is also pointed out there, the notion of refining needed for the definition of the continuum limit should incorporate this physical vacuum. In particular, \cite{refining} suggests to use Pachner moves \cite{Pachner}, which also implement the dynamics of the system \cite{hoehn}, for refining the state. We will see that this idea will indeed turn out to be essential.

A further motivation is the notion of duality. While the AL vacuum is peaked on fluxes and has maximal uncertainty in the holonomies, the new (BF) vacuum which we introduce is peaked on flat holonomies and will have maximal uncertainty in certain flux observables. A description and interpretation of the classical phase space for LQG, based on using the constraints of BF theory as gauge fixings, has been given in \cite{FGZ} (see also the related discussions in \cite{bobienski-bianchi,bahr}). The notion of duality leads also to a flux representation of LQG \cite{flux}, which is dual to the more common holonomy representation. Indeed, it was pointed out in \cite{flux3} that in order to construct a flux representation in complete analogy  with the holonomy representation, one  needs to replace the AL vacuum  by a ``dual'' BF vacuum.

The setup in this work will be a simplicial version of LQG (see also \cite{QSD7}). This is needed in order to implement Pachner moves as refinement. To make the notion of a BF vacuum for LQG concrete, we have to ($a$) define the vacuum state, and ($b$) define a notion of refinement. For point ($c$), we first have to specify a cylindrically consistent observable algebra (i.e. one that is consistent with the notion ($b$) of refinement). This latter will be different from the standard holonomy--flux algebra of LQG, which allows us to evade the uniqueness theorems. We will then show that the Hilbert space of gauge--invariant functions can be generated from the new vacuum by (exponentiated) flux observables.

We will restrict ourselves to $(2+1)$--dimensional gravity, for which the BF state is indeed the physical vacuum, and hence implement the ideas of \cite{refining}. The extension to the $(3+1)$--dimensional case is developed in \cite{to appear}.

\section*{2. Setup}

Here we sketch the set--up of our construction. We consider a two--dimensional orientable smooth manifold, together with an atlas of coordinate charts and an auxiliary metric. To this manifold, we associate a set of embedded triangulations $\Delta^*$, their dual triangulations $\Delta$, and denote by $\Gamma$ the one--skeleton of $\Delta$. Each dual triangulation consists of (three--valent) vertices $v$, oriented edges $e$, and faces. By embedded, we simply mean that the vertices carry coordinates. We assume that the edges of the triangulation (which we will denote by $e^*$) are geodesics with respect to the auxiliary metric. To this end we also assume that the triangulation is sufficiently fine in order to ensure that the geodesics are well--defined.

\begin{center}
\begin{figure}[h]
\includegraphics[scale=0.45]{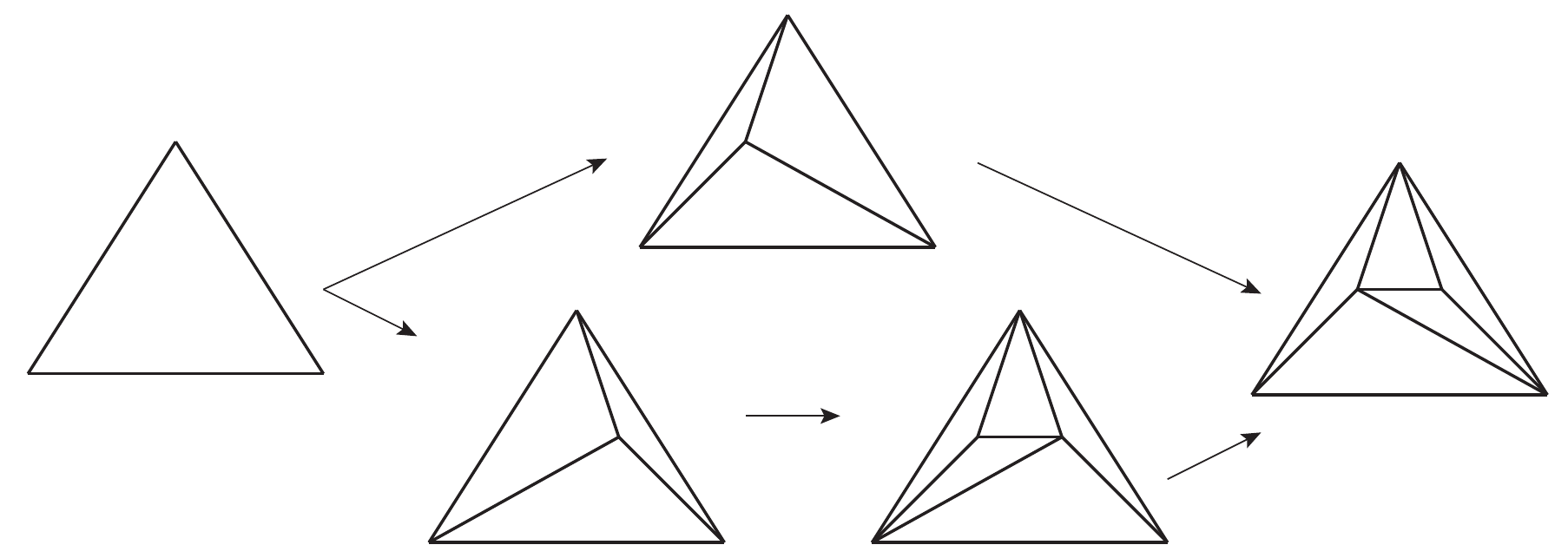}
\caption{Example of refinement via 1--3 and 2--2 moves. The rightmost triangulation is finer than the leftmost one, and represents the common refinement $\overline{\Delta}$ of the triangulations on the top and on the bottom.}
\label{fig:refined}
\end{figure}
\end{center}

We endow the set of triangulations with a partial order denoted by $\prec$. A (dual) triangulation $\Delta'$ is said to be finer than a (dual) triangulation $\Delta$ if $\Delta'$ can be obtained from $\Delta$ by 1--3 and 2--2 Pachner moves \cite{Pachner}, as shown for example on figure \ref{fig:refined}. For the 1--3 move, one needs to specify in addition a set coordinates for the new node of the triangulation. We will see that freedom in this choice is essential for realizing a notion of spatial diffeomorphism invariance with respect to the embedding.

Two triangulations $\Delta$ and $\Delta'$ can be compared if they have a common refinement $\overline{\Delta}$. An example is depicted in figure \ref{fig:refined}, showing that two triangulations obtained via a 1--3 move, but each with a different placement of the new vertex, have a common refinement.

Concerning the group--theoretic data, we associate to each edge a space $\mathcal{F}(G)$ of functions over the group $G$. The topology of this space will be specified later on. The group $G$ in question can be finite or a compact semi--simple Lie group, and either Abelian or non--Abelian.

The notion of refinement that we are going to use is based on Pachner moves. Pachner moves for triangulated two-dimensional surfaces consist of the so--called 1--3, 3--1, and 2--2 moves, and two such triangulated surfaces are PL--homeomorphic if one can be transformed into the other by a finite number of these moves. Since here we are interested in refining operations, we will consider only the 1--3 as well as the 2--2 Pachner moves (i.e. discard the 3--1 moves). This is sufficient to ensure that topologically equivalent triangulations have common refinements. The notion of geometric triangulations includes embedding information for the vertices, and at least for the planar case it follows from \cite{geometric-trian} that any two such triangulations have a common refinement. 

Our task is to find the observables that are cylindrically consistent, i.e. commuting with the refinement operations. We will first focus on closed holonomies and then on the integrated simplicial fluxes.

\section*{3. Cylindrical consistency of closed holonomies} 

The cylindrical consistency of closed holonomies follows from the definition of the refining Pachner moves in the holonomy representation, which we are now going to present. 

In the holonomy representation, the action of the Pachner moves can directly be deduced from the geometrical interpretation of BF theory as describing flat connections. Let $h_\gamma$ denote the holonomy along a (closed) path $\gamma$. Gluing a tetrahedron to the surface might locally change this path to $\gamma'$. However, since this amounts to adding only pieces of flat holonomies, the deformation of the path will not change the holonomy, and one can write that $h_\gamma=h_{\gamma'}$.

This requirement determines the action of the Pachner moves uniquely. From this it will follow that closed holonomy observables are cylindrically consistent, i.e. commuting with the action of refining Pachner moves.

\paragraph*{\bf 1--3 Pachner moves.}

\begin{center}
\begin{figure}[h]
\includegraphics[scale=0.45]{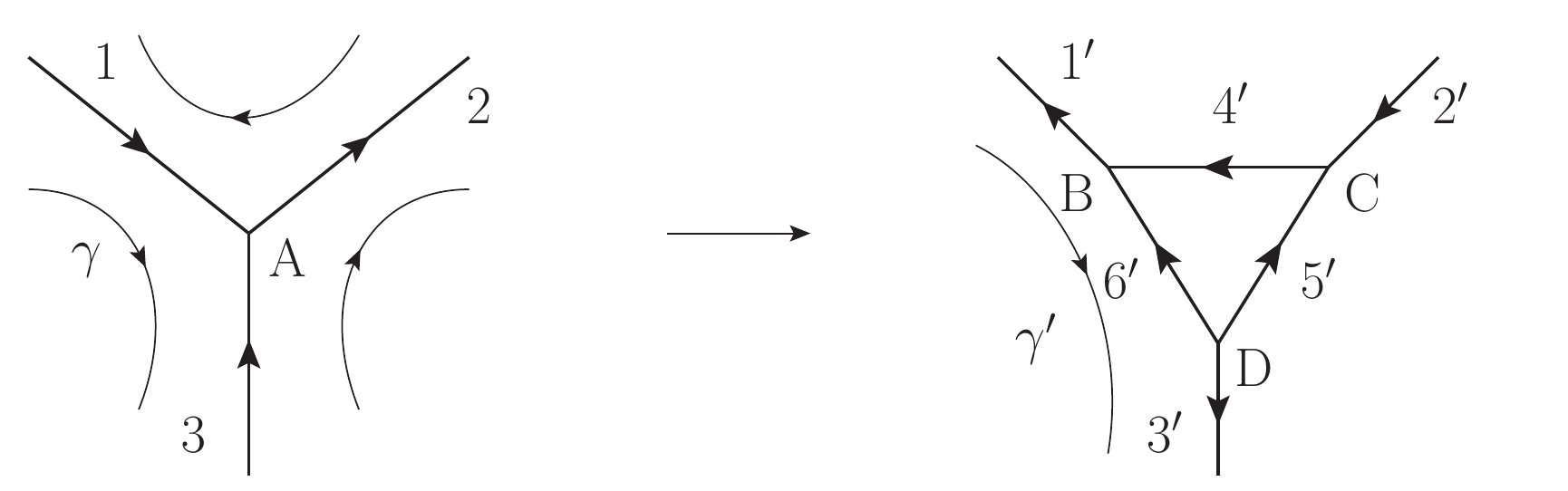}
\caption{Local change of the graph $\Gamma$ to $\Gamma'$ under the action of a 1--3 move. Notice the reversal of the edges, which is due to the fact that the 1--3 move glues a tetrahedron onto a triangle, so that one actually replaces oppositely oriented half--edges in the dual. The path $\gamma$ is changed to $\gamma'$.}
\label{fig:1-3}
\end{figure}
\end{center}

Let us consider a graph $\Gamma$ consisting of a vertex of $\Delta$ labeled $v_\A$ and its three edges\footnote{Orientations in $\Gamma$ can always be adjusted, and under the change of orientation of an edge $e$ we have $\psi(\dots,g_e,\dots)\mapsto\psi(\dots,g_{e^{-1}},\dots)=\psi(\dots,g_e^{-1},\dots)$.} $e_1$, $e_2$ and $e_3$. Gluing a tetrahedron to the triangulation changes the graph $\Gamma$ to $\Gamma'$ as depicted on figure \ref{fig:1-3}. The vertex $v_\A$ is replaced by three vertices, $v_\B$, $v_\C$, and $v_\D$, and three new edges are introduced.

We now require that the holonomies stay the same, i.e. that $h_\gamma=h_{\gamma'}$, where $\gamma$ is a path in $\Gamma$, and $\gamma'$ the path in $\Gamma'$ that goes along the new edges but stays in the same face. To this end, we have to define $\gamma'$ as a function of the path $\gamma$. We choose the rule that $\gamma'$ goes through the same vertices and in the same order as $\gamma$. Since the 1--3 move replaces the vertex $v_\A$ with three new vertices, the passage through $v_\A$ can be replaced by any combination of paths through the three new vertices. Therefore, for the holonomies to stay the same, we need the following replacement rules (we use a counter--clockwise orientation of the faces):
\ba\label{holo1}
&\gm{3}\g{1}\rightarrow\g{3'}\gm{6'}\gm{1'},\qquad\qquad
\gm{1}\gm{2}\rightarrow\g{1'}\g{4'}\g{2'},&\nn \\
&\g{2}\g{3}\rightarrow\gm{2'}\g{5'}\gm{3'}.&
\ea
This implies that the holonomy around the new face is trivial, i.e. $\gm{6'}\g{4'}\g{5'}=\mathbb{I}$. Implementing the conditions $h_\gamma=h_{\gamma'}$ given by the replacements \eqref{holo1} with delta functions, we can define the wave function $\psi'$ after the move to be
\ba\label{holo2}
&&\psi'(\g{1'},\dots,\g{6'},\dots)\nn\\
&=&\int\delta(\gm{1}\g{3}\g{3'}\gm{6'}\gm{1'})\delta(\g{2}\g{1}\g{1'}\g{4'}\g{2'})\\
&&\phantom{\int}\delta(\gm{3}\gm{2}\gm{2'}\g{5'}\gm{3'})\psi(\g{1},\g{2},\g{3},\dots)\ \bd\g{1}\bd\g{2}\bd\g{3},\nn
\ea
where $\delta(\cdot)$ is the group delta function and $\bd{g}$ is the normalized Haar measure (a counting measure in the case of finite groups).

We can solve the first two delta functions for $\g{2}$ and $\g{3}$, which leads to the replacement of the third delta function by $\delta(\gm{6'}\g{4'}\g{5'})$, thereby implementing that the holonomy around the new face of $\Delta$ is trivial. We are left with an integral over $\g{1}$,
\ba\label{holo3}
&&\psi'(\g{1'},\dots,\g{6'},\dots)\\
&=&\!\!\int\delta(\gm{6'}\g{4'}\g{5'})\psi(\g{1},\g{2},\g{3},\dots)_{\scriptsize{\left|
\begin{array}{l}
\!\!\g{2}=\gm{2'}\gm{4'}\gm{1'}\gm{1}\\
\!\!\g{3}=\g{1}\g{1'}\g{6'}\gm{3'}\end{array}
\right.
}}\!\bd\g{1}.\nn
\ea
This integral over $\g{1}$ amounts to an averaging over the gauge group at the vertex $v_\B$ (one will obtain one of the other two vertices if one solves \eqref{holo2} for other variables). Assuming that the wave function $\psi$ is gauge--invariant at $v_A$, we can therefore gauge fix $\g{1}$ to $\g{1}=\gm{1'}$, and find that the action of the 1--3 Pachner move on a gauge--invariant wave function is given by
\ba\label{holo4}
&&\psi'(\g{1'},\dots,\g{6'},\dots)\\
&=&\delta(\gm{6'}\g{4'}\g{5'})\psi(\g{1},\g{2},\g{3},\dots)_{\scriptsize{\left|
\begin{array}{l}
\!\!\g{1}=\gm{1'}\\
\!\!\g{2}=\gm{2'}\gm{4'}\\
\!\!\g{3}=\g{6'}\gm{3'}\end{array}
\right.
}}.\nn
\ea

One can see that the Pachner move has added three new edges, and hence three new holonomy variables. However, these new variables are restricted by a delta function, and there are in addition two gauge degrees of freedom. Therefore, no new ``true'' degree of freedom was added.

Later on we will need to choose a vertex $v$ to be the root $r$ of a tree, thereby fixing a reference frame. In the case where the root is $r=v_\A$, we have to choose one of the three new vertices as the new root $r'$. With the gauge fixing leading to the expression \eqref{holo4}, we have $r'=v_\B$.

By construction, the action of the 1--3 Pachner moves commutes with the action of (closed) holonomy operators. These just act as multiplication operators, and are gauge--invariant if the path is closed.\\

\paragraph*{\bf 2--2 Pachner moves.}

\begin{center}
\begin{figure}[h]
\includegraphics[scale=0.45]{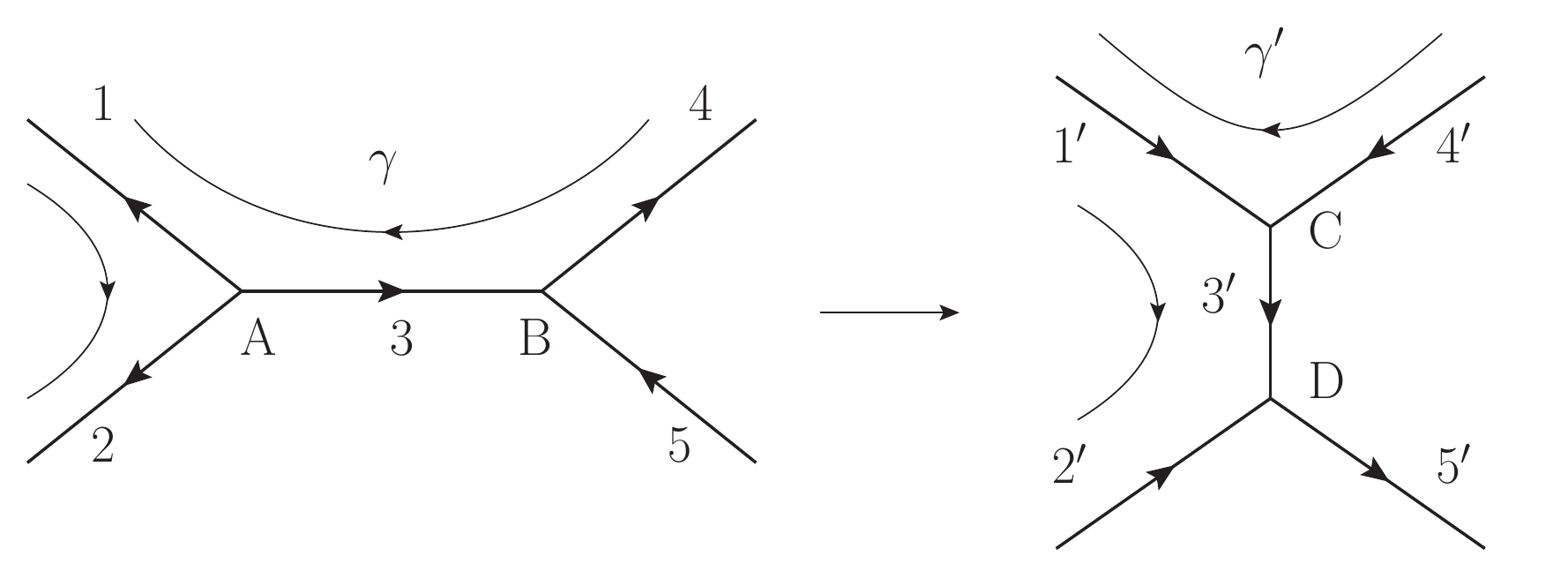}
\caption{Local change of the graph $\Gamma$ to $\Gamma'$ under the action of a 2--2 move, with the example of a path $\gamma$ changed to $\gamma'$.}
\label{fig:2-2}
\end{figure}
\end{center}

The study of the 2--2 Pachner move (represented on figure \ref{fig:2-2}) is similar. We adopt the following replacement rule for a path $\gamma\rightarrow\gamma'$. Let $\gamma$ be a path starting along the edge $e_i$, going possibly through $e_3$, and ending along $e_j$, with $i,j=1,2,4,5$. Then we define $\gamma'$ in an obvious way as passing through the corresponding replaced edges $e_{i'}$.

This rule is described by the following replacements of the path holonomies:
\ba\label{holo5}
&&\g{2}\gm{1}\rightarrow\gm{2'}\g{3'}\g{1'},\qquad
\g{1}\gm{3}\gm{4}\rightarrow\gm{1'}\g{4'},\nn\\
&&\gm{5}\g{3}\gm{2}\rightarrow\g{5'}\g{2'},\qquad
\g{4}\g{5}\rightarrow\gm{4'}\gm{3'}\gm{5'}.
\ea
The last replacement rule in \eqref{holo5} is redundant since it follows from the first three ones. For a gauge--invariant wave function we can again use a gauge fixing. Since there are two vertices, $v_\A$ and $v_\B$, we can gauge fix two variables, e.g. $\g{2}=\gm{2'}$ and $\g{3}=\mathbb{I}$. Then the new wave function $\psi'$ after the 2--2 Pachner move becomes
\ba\label{holo6}
&&\psi'(\g{1'},\dots,\g{5'},\dots)\\
&=&\psi(\g{1},\dots,\g{5},\dots)_{\scriptsize{\left|
\begin{array}{l}
\!\!\g{1}=\gm{1'}\gm{3'}\\
\!\!\g{2}=\gm{2'}\\
\!\!\g{3}=\mathbb{I}
\end{array}
\right.
\left|
\begin{array}{l}
\!\!\g{4}=\gm{4'}\gm{3'}\\
\!\!\g{5}=\gm{5'}
\end{array}
\right.
}}\nn.
\ea

By construction, the action of the 2--2 Pachner moves does therefore commute with closed holonomy operators.

If $r=v_\A$ or $r=v_\B$ is a root vertex, we need to choose, in agreement with the gauge fixing in \eqref{holo6}, $r'=v_\D$ as the new root.

Note that one can use the Peter--Weyl transform to write the Pachner moves in the holonomy representation into the spin representation, in which the moves appear as the gluing to the hypersurface of Ponzano--Regge \cite{PR} amplitudes corresponding to a tetrahedron. The holonomy operators appear as a so--called tent move in the spin representation \cite{bonzomRecurrence}, and the closed holonomies and Pachner moves commute due to the Biedenharn--Elliot identity \cite{to appear}.

\section*{4. Cylindrical consistency of integrated\\ simplicial flux operators}

Let us now discuss the cylindrical consistency of the flux operators. First, not that the fluxes are elements of the Lie algebra, and as such act as derivative operators (the left or right multiplication on the group can be used to establish an isomorphism between the Lie algebra and vector fields). It may therefore seem that $G$ is required to be a Lie group in order for the construction to make sense. However, we are going to use exponentiated flux operators, which act as shifts on the group variables, and can hence also be defined for finite groups.

Since we are considering a triangulation and its dual graph $\Gamma$, we will be interested in the so--called ``simplicial'' or ``geometrical'' fluxes \cite{viqar,QSD7,flux,FGZ}, as opposed to the standard fluxes of LQG \cite{books}. Let us define these simplicial fluxes in the case where $G=\SU(2)$ and in $(2+1)$ dimensions. Let $e^*\in\Delta^*$ be the oriented edge dual to $e\in\Delta$. We assume that the pair $(e,e^*)$ is positively oriented. Given the continuum triad field $E=E^i_a\tau_i\bd x^a$ (where $\tau_i$ is a basis of $\mathfrak{su}(2)$ and $a=1,2$ a spatial one--form index), the fluxes are defined by
\be\label{flux1}
X_e=\int_{e^*}\h{e^*(t),e(0)}E_a\big(e^*(t)\big)(\dot{e}^*)^a(t)\hm{e^*(t),e(0)}\ \bd t,
\ee
where $t\in[0,1]$ is a parametrization of the edge $e^*$ such that $e$ and $e^*$ intersect at $t=1/2$, and $h_{e^*(t),e(0)}$ is the parallel transport from the point $e^*(t)$ to the source vertex $e(0)$ of $e$. This holonomy starts at $t\in e^*$, goes along $e^*$ until it reaches the point $e\cap e^*$, and then goes along $e$ until its source $e(0)$.

The advantage of using these fluxes is that under a gauge transformation or an orientation reversal of the edge they transform as
\be\label{flux properties}
X_e\mapsto g(e(0))X_e\big(g(e(0)\big)^{-1},\quad X_{e^{-1}}=-\g{e}X_e\gm{e}.
\ee
These properties and definitions can be generalized to other (Lie) groups, such as $\mathrm{SL}(2,\mathbb{C})$ \cite{wieland}.

The action of the fluxes leaves the space of functions $\psi:G^{|E|}\rightarrow\mathbb{C}$ over a fixed dual graph (with $|E|$ edges) invariant. This action is defined as
\be
X_e^i\triangleright\psi=\mathrm{i}\hbar\big(R^i_e\psi\big),
\ee
where $R^i_e$ is the left invariant vector field acting on functions $\psi$ on the group as the right--derivative
\be
\big(R^i\psi\big)(g)=\left.\frac{\bd}{\bd t}\psi\Big(ge^{tT^i}\Big)\right|_{t=0}.
\ee
Here $\{T^i\}$ is a choice of basis of generators of the Lie group $G$, and we have used the left multiplication to identify the Lie algebra and the derivative operators.

Geometrically, a flux variable expresses the vector describing the edge $e^*$ in the reference frame given by the source vertex of $e$. Since a 2--2 move replaces an edge of $\Delta^*$ by a transversal edge, we see that a flux variable in itself cannot be cylindrically consistent.

The way around this issue is to introduce ``integrated'' flux observables $\bX_{\pi^*}$ associated to paths $\pi^*$ in the one--skeleton $\Gamma^*$ of the triangulation $\Delta^*$ \cite{freidel-louapre}. Essentially, these flux observables are defined as the sum of the individual fluxes $X_e$ associated to the various edges $e$ dual to the elements $e^*$ constituting the path $\pi^*$, and possibly inverted in order to have the same orientation. In order to make this construction consistent and well--defined, the individual fluxes have to be transported (with the adjoint action) to a common reference point, which we define as the source $\pi(0)$ of the path $\pi$ (i.e. the source of the first edge of $\pi$). The integrated flux observables are therefore defined as
\be\label{flux observable}
\bX_{\pi^*}:=\sum_{e\subset\pi}\h{e(0),\pi(0)}X_e\hm{e(0),\pi(0)},
\ee
where $h_{e(0),\pi(0)}$ is the holonomy going from the source of the edge $e$ to the reference point $\pi(0)$, and the fluxes have to be inverted if their orientation does not agree with the one outgoing from $\pi(0)$ that defines the general orientation of the path. In the Abelian case the definition \eqref{flux observable} reduces simply to the sum of the fluxes, without the adjoint action.

This integrated flux observable corresponds geometrically to the displacement vector between the nodes $\pi^*(0)$ and $\pi^*(1)$ of the triangulation. Under the action of a gauge transformation, it transforms by the adjoint action of the group in the frame in which the individual fluxes are transported, i.e. the source vertex $\pi(0)$.

The composition of these flux observables is defined in the standard way as follows. Suppose that $\bX_{\pi_1^*}$ is a flux observable defined along a path $\pi_1^*$ and $\bX_{\pi_2^*}$ is defined along $\pi_2^*$ such that $\pi_1(1)=\pi_2(0)$. Transporting the fluxes in the same frame in order to add them, we have the composition rule
\be
\bX_{\pi_2^*}\circ\bX_{\pi_1^*}=\bX_{\pi_1^*}+h_{\pi_1}^{-1}\bX_{\pi_2^*}h_{\pi_1}^{\vphantom{-1}},
\ee
where $h_{\pi_1}$ is the holonomy from $\pi_1(0)$ to $\pi_2(0)$.

\begin{center}
\begin{figure}[h]
\includegraphics[scale=0.45]{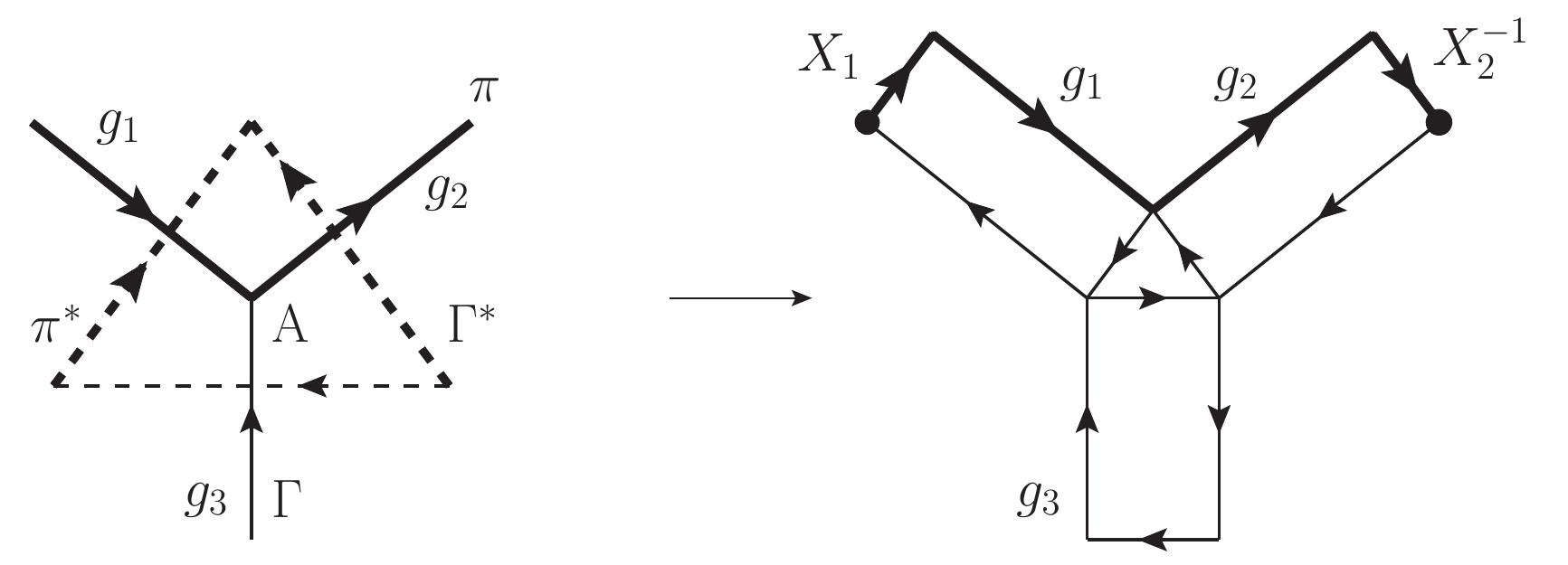}
\caption{Replacement of a portion of triangulation (dashed triangle) and its dual by a ribbon graph that contains both the holonomies and the fluxes. Our convention is such that the ribbons have a clockwise orientation, and that the pairs $(\text{holonomy},\text{flux})$ have positive orientation. A path defining an integrated flux $\bX_{\pi^*}$ is represented with thick lines.}
\label{fig:ribbon}
\end{figure}
\end{center}

It is convenient to introduce the ribbon picture \cite{ribbons}, which simultaneously represents the graph $\Gamma$ and its dual $\Gamma^*$ as depicted in figure \ref{fig:ribbon}. A ribbon can be assigned holonomy and flux data with the composition rule $(g_1,0)\circ(\mathbb{I},X_1)=(g_1,X_1)$. Ribbons can then be composed following a path $\pi^*$. This defines the integrated fluxes through
\be
(g_2,X_2^{-1})\circ(g_1,X_1)=\big(g_2g_1,\bX_{\pi^*}=X_1+g_1^{-1}X_2^{-1}g_1^{\vphantom{-1}}\big),
\ee
which is a semi--direct product structure.

Finally, one can use the fact that the Poisson bracket between the simplicial fluxes \eqref{flux1} and the holonomies reproduce the Poisson structure on $T^*G$ to derive the Poisson action of the integrated flux \eqref{flux observable} on a function $\psi(g_e)$. If $e\not\subset\pi$, this action is vanishing. If $e\subset\pi$, it is given by
\be
\big\{\bX_\pi,\psi(g_e)\big\}=\Big(\Big[\hm{\pi(0),e(0)}R\h{\pi(0),e(0)}\Big]^i\psi\Big)(g_e).
\ee
%Note that the parallel transport in this definition does not involve the edge $e$ since it stops at the source $e(0)$.

Since the Pachner moves that we are considering are only refinement moves, they cannot remove the reference frame $\pi(0)$ of an integrated flux observable. These flux observables are cylindrically consistent. Indeed, one can understand heuristically that the Pachner moves add only pieces of flat geometry, for which the Gauss constraints are satisfied. Thus the vector associated to the path $\pi^*$ does not change under a Pachner move.

In order to show this explicitly, we have once again to specify the replacement rules and argue that the observables commute with the Pachner moves.

For the 1--3 Pachner move, the invariance of the integrated fluxes is immediate since the edges $e^*$ of the triangulation defining the path $\pi^*$ cannot be affected by the move. However, in order to show this explicitly, one has to keep track of the new holonomies that are introduced by the 1--3 move. Our replacement rule is such that the new path in the ribbon graph after the 1--3 move has to start from the same source $\pi(0)$, end at the same terminal point $\pi(1)$, go along the same flux variables (or their inverse), and transport these fluxes using the new holonomies.

\begin{center}
\begin{figure}[h]
\includegraphics[scale=0.45]{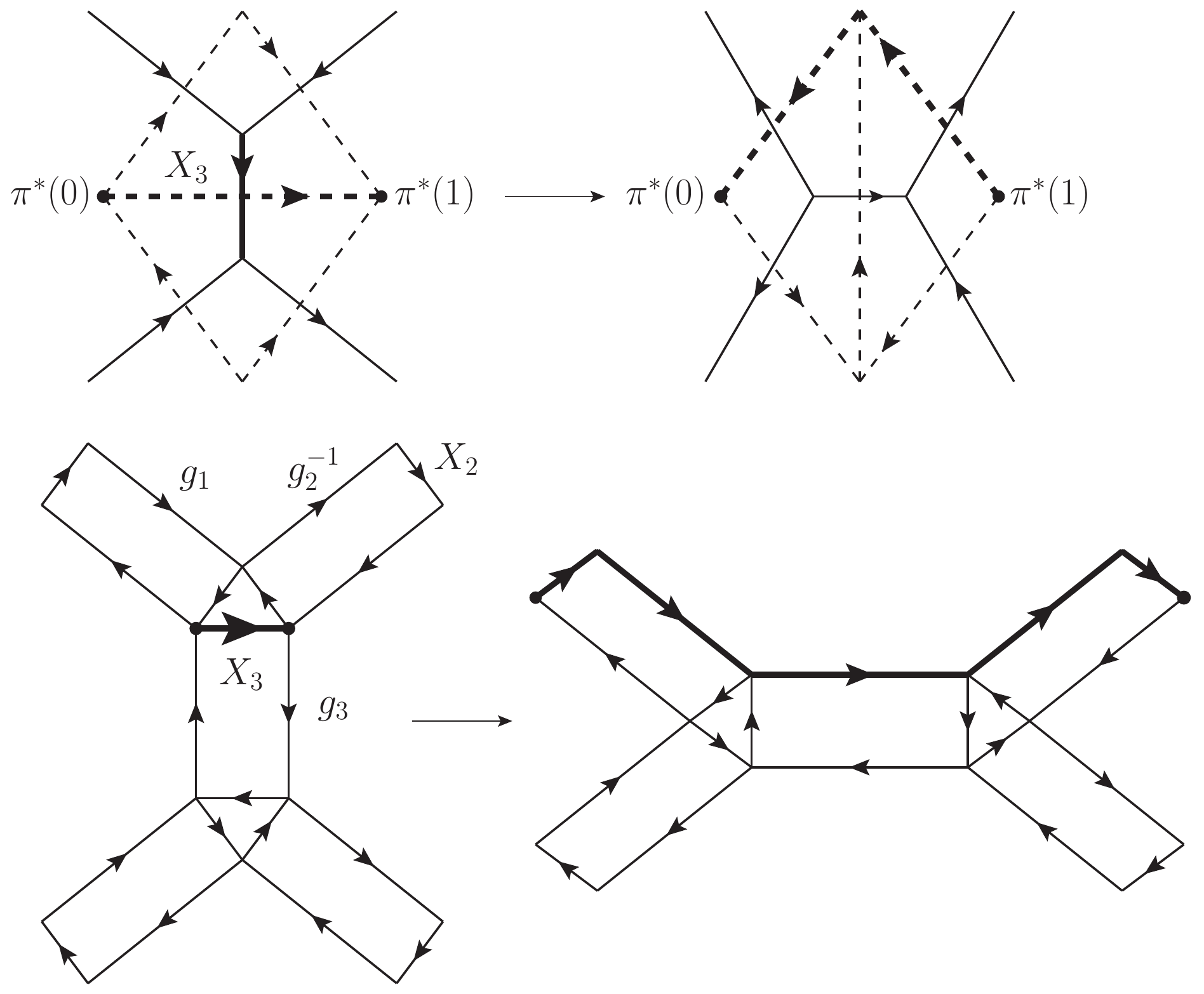}
\caption{Replacement under a 2--2 move of an integrated flux by another flux that defines the same displacement vector in the triangulation but goes along a different path.}
\label{fig:2-2Flux}
\end{figure}
\end{center}

Concerning the 2--2 move, if the path $\pi^*$ does not go along the internal edge of the triangulation that is being flipped by the move, the invariance of the integrated flux follows from the same reasoning as in the 1--3 case discussed above. If the path (or a portion of it) corresponds to the edge that is being flipped, one has to use the replacement rule depicted on figure \ref{fig:2-2Flux}. This replacement rule is unique as long as one follows the arrows of the ribbon representation.

Finally, let us conclude with an important remark concerning the reference frame to which the individual fluxes are transported. It can happen that this reference frame coincides with a vertex of $\Gamma$ that is moved by the 1--3 or 2--2 move. This is illustrated for example on figure \ref{fig:1-3}, where the vertex $v_\A$ is replaced by the three vertices $v_\B$, $v_\C$, and $v_\D$. Therefore, if the reference frame of the path defining $\bX_{\pi^*}$ coincides with the vertex $v_\A$, there is an ambiguity in the choice of placement of the reference frame after the Pachner move. While this is not problematic for the holonomy observables because of their gauge invariance, it affects the integrated fluxes since they are gauge--covariant. Since the construction of the measure in the next section will involve a choice of a maximal tree with root $r$, we can choose to define the integrated fluxes transported to the root.  We discussed how the root behaves under Pachner moves in section 3.

Gauge--invariant observables can be easily obtained by taking the Lie algebra trace over two integrated fluxes transported to a common reference frame, or by group averaging combinations of exponentiated fluxes as in the next section. Details on the Poisson algebra of these fluxes and holonomies will be discussed in \cite{to appear}.

\section*{5. Measure} 

The AL measure can be characterized by the evaluations of the positive linear functional $\mu_\text{AL}$ on a spin network basis, which itself can be generated by the application of holonomy observables $\psi_{\{j\}}(\{g\})$ to the AL vacuum state $\eta_\text{AL}(\{g_e\})\equiv1$. The holonomies $\psi_{\{j\}}(\{g\})$, with $\{j\}$ denoting the set of representations labeling the edges, are multiplication operators and lead to the spin network basis. More precisely, the functional $\mu_\text{AL}$ is defined on this basis as $\mu_\text{AL}(\psi_{\{j\}})=\delta_{\varnothing,\{j\}}$, which is non--vanishing if and only if all representation labels are trivial.

To construct the new measure we will proceed similarly, but however dualizing every ingredient. Instead of a constant function in the holonomy representation, we consider a constant\footnote{This applies in the gauge--fixed version. In the gauge--variant version, the constant function is replaced by a function which has only support on the solution of the Gauss constraints and is otherwise constant.} function in the flux representation. Furthermore, the spin network basis generated by holonomies is replaced by a dual basis, which is generated by exponentiated (integrated) flux observables. All this is easier to consider with a choice of gauge fixing. We will also consider for now a fixed triangulation and its dual triangulation.

A gauge fixing can be introduced by choosing a maximal tree $\mathcal{T}$ in $\Gamma$ with root $r$. Group elements associated to edges $t$ of the tree can be gauge--fixed to the identity. Edges which are not included in the tree are called leaves $\ell$ and are in one--to--one correspondence with the fundamental cycles of $\Gamma$. We define the cycle $c_\ell$ associated to the leave $\ell$ by choosing the same starting vertex and orientation for the cycle as for the leave (edge) $\ell$. Apart from $\ell$, all other edges in the cycle are elements $t$ of the tree. We will  denote the holonomy associate to the cycle $c_\ell$ by $\mathcal{C}_\ell$. A BF vacuum state that is peaked on a locally and globally flat connection for $\Gamma$ is given in the holonomy representation by $\eta_\text{BF}=\prod_\ell\delta(\mathcal{C}_\ell)\ \dot{=}\ \prod_\ell\delta(g_\ell)$, where $\dot{=}$ denotes the gauge--fixed expression and $g_\ell$ is the group element associated to the leave $\ell$.

We can now consider the exponentiated integrated flux observables associated to the leaves $\ell$ and transported to the root $r$. These act as right translations by group elements $\text{Ad}_{t_\ell}(h_\ell)$. We will therefore denote the exponentiated flux observables by $R_{\{ \text{Ad}_{t_\ell}(h_\ell)\}}$, where $t_\ell$ is the holonomy associated to the unique path going from the source vertex of the leave $\ell$ to the root $r$, and we have defined $\text{Ad}_g(h)=ghg^{-1}$. The action on the vacuum is 
\be
R_{\{\text{Ad}_{t_\ell}(h_\ell)\}}\eta_\text{BF}\ \dot{=}\ \prod_\ell\delta (g_\ell h_\ell).
\ee
For non--Abelian groups, this results in general in a gauge--covariant function at the root $r$. For Abelian groups, $R_{\{h_\ell\}}\eta_\text{BF}$ can be understood as the dual of the (gauge--invariant) spin network basis, labelled by group elements $\{h_\ell\}$ instead of representation labels.\\

\paragraph*{\bf Abelian groups.}

To define a measure in analogy with the AL one, let us switch to the ``spin representation''. First, we restrict to Abelian groups and formally define the duality between group and spin representation by $\langle g|j\rangle=\chi_j(g)$, with $\chi_j$ a character of $G$ and $j$ an element of the Pontryagin dual. For the gauge--fixed functions, the change to the spin representation leads to functions $\widetilde\psi(\{j_\ell\})$, the vacuum is represented by $\widetilde\eta_\text{BF}\ \dot{\equiv}\ 1$, and the right translations act as multiplication operators in the following way:
\be
R_{\{h_\ell\}}\widetilde\psi(\{j_\ell\})\ \dot{=}\,\left(\prod_\ell\chi_{j_\ell}(h_\ell)\right)\widetilde\psi(\{j_\ell\}).
\ee
Therefore, $\widetilde\chi_{\{h_\ell\}}(\{j_\ell\}):=\prod_\ell\chi_{j_\ell}(h_\ell)$ is a (possibly non--normalizable) basis for the space of functions on the space of fluxes, with the property $\widetilde\chi_{\{h_\ell\}}(\{j_\ell\})\widetilde\chi_{\{h'_\ell\}}(\{j_\ell\})=\widetilde\chi_{\{h_\ell h'_\ell\}}(\{j_\ell\})$. We can now define a measure on this space of functions by
\be\label{measureA}
\mu_\text{BF}(\widetilde\chi_{\{h_\ell\}})=\prod_\ell\tilde\delta(h_\ell). 
\ee
For finite groups, $\tilde\delta$ is defined to be the Kronecker symbol. For Lie groups, we have two possibilities for specifying $\tilde\delta$. First, we can choose $\tilde\delta$ to coincide with the group delta function. As we will show below, in this case the resulting inner product coincides with the one of square integrable functions $L_2\big(G^{|E|},\bd^{|E|}g\big)$ with the Haar measure on $G$. Another possibility is to formally define $\tilde\delta$ to be $\tilde\delta(h)=1$ if $h=\mathbb{I}$ and to be vanishing otherwise. For $G=\text{U}(1)$, this leads to a Bohr compactification of the dual $\mathbb{Z}$ to $\text{U}(1)$ \cite{Besicovitch}, which turns the vacuum into a normalizable state.
 
To prove the above statement, we compute the $L_2\big(G^{|E|},\bd^{|E|}g\big)$ inner product between two states $R_{\{h_\ell\}}\eta_\text{BF}$ and $R_{\{h'_\ell\}}\eta_\text{BF}$ which via the functional \eqref{measureA} is given by $\prod_\ell\tilde\delta\big({h'}_\ell^{-1}h_\ell\big)$:
\ba
&&\int\overline{R_{\{h_\ell\}}\eta_\text{BF}}\ R_{\{h'_\ell\}}\eta_\text{BF}\ \bd^{|E|}g_e\nn\\
&=&\int\overline{R_{\{ h'^{-1}_\ell h_\ell\}}\prod_\ell\delta(g_\ell)}\ \prod_{\ell'}\delta(g_{\ell'})\ \bd^{|L|}g_\ell\\
&=&\prod_\ell\delta\big({ h'}^{-1}_\ell h_\ell\big),\nn
\ea
where we applied the gauge fixing to go from the first to the second line, and $|L|$ denotes the number of leaves. Thus, if the $\tilde\delta$ are chosen to be the group delta functions, the inner product defined via the functional \eqref{measureA} coincides on a fixed graph with the inner product defined via the Haar measure.\\

\paragraph*{\bf Non--Abelian groups.}

The case for non--Abelian groups works in a similar way. However, in contrast to the Abelian case, the space of gauge--invariant functions is not parametrized by $G^{|L|}$ but by $G^{|L|}/\text{Ad}_G$, where $\text{Ad}_G$ denotes the remaining action of the gauge group at the root $r$ on the space of gauge--fixed functions. We now have shift operators $R_{\{\text{Ad}_{t_\ell}(h_\ell)\}}$ which with the BF vacuum defined as before lead to basis states
\be
\widetilde\chi_{\{h_\ell\}}:=R_{\{\text{Ad}_{t_\ell}(h_\ell)\}}\eta_\text{BF}
\ee
that are gauge--variant at the root $r$. This can be cured by a group averaging, leading to gauge--invariant states
\be
\mathcal{G}(\widetilde\chi_{\{h_\ell\}}):=\int R_{\{ \text{Ad}_{t_\ell u}(h_\ell)\}}\eta_\text{BF}\ \bd u,
\ee
where $u\in G$ is the group averaging parameter.

To reproduce the inner product of $L_2(G^{|E|},\bd^{|E|}g)$, we have to define the product between two such states to be
\be
\langle\mathcal{G}(\widetilde\chi_{\{h_\ell\}})\,,\,\mathcal{G}(\widetilde\chi_{\{h'_\ell\}})\rangle:=\int\prod_\ell\delta\big({h'}^{-1}_\ell uh_\ell u^{-1}\big)\ \bd u.
\ee
A measure leading to this inner product, i.e. satisfying $\langle\mathcal{G}(\widetilde\chi_{\{h_\ell\}})\,,\,\mathcal{G}(\widetilde\chi_{\{h'_\ell\}})\rangle=\mu_\text{BF}\big(\overline{\mathcal{G}(\widetilde\chi_{\{h_\ell\}})}\mathcal{G}(\widetilde\chi_{\{h'_\ell\}})\big)$ is given by
\be\label{measureB}
\mu_\text{BF}\big(\mathcal{G}(\widetilde\chi_{\{h_\ell\}})\big)=\prod_\ell\delta(h_\ell).
\ee

As with the Abelian groups, one could also consider some compactification of the dual of $G$, or rather of the dual of the maximal torus of $G$. We leave this question for future explorations. This construction of the measure is independent of the choice of tree as long as the delta functions on the right hand side of \eqref{measureA} and \eqref{measureB} are with respect to group translation invariant measures.
 
\section*{6. On the projective limit and\\ spatial diffeomorphisms} 
 
So far we discussed the inner product on a fixed triangulation. To compare two states on two different triangulations, we need to consider a common refinement of these triangulations\footnote{Thus, the set of triangulations which we consider should be directed, i.e. for any pair of triangulations there should exist a common refinement. This is for instance the case for geometric triangulations \cite{geometric-trian}.}. To this end, the measure or inner product has to be cylindrically consistent, so that it does not depend on the precise choice of common refinement.

This is the case here since the construction of the inner product only involves cylindrically consistent observables. Thus, one can expect independence of the choice of refinement. However, if we choose in \eqref{measureA} and \eqref{measureB} the group delta functions on the right hand side, we see that 1--3 moves will lead to additional factors of $\delta(\mathbb{I})$ for the measure and the inner product. To cure these divergencies for the inner product we can choose a (heat kernel) regularization for the delta functions, and divide the inner product of two states by the norm of a reference state (in this case the BF vacuum). That is, we define a modified inner product as (see also \cite{bahr})
\ba
\langle\psi_1,\psi_2\rangle'=\lim_{\varepsilon\rightarrow0}\frac{\langle\psi_1,\psi_2\rangle_\varepsilon}{\langle\eta_\text{BF},\eta_\text{BF}\rangle_\varepsilon},
\ea
where $\varepsilon$ indicates the regulator.
  
Let us also mention a notion of spatial diffeomorphism invariance for the vacuum. Consider a triangle subdivided by a 1--3 move but with the inner vertex placed at two different positions leading to two states $\psi_1$ and $\psi_2$. Figure \ref{fig:refined} shows that there exists a common refinement and indeed the two different ways to obtain this refinement starting with the triangle via $\psi_1$ and $\psi_2$ lead to the same state. Thus the inner product identifies the two states which differ by a vertex translation. This nicely reflects the notion of vertex translations as a diffeomorphism symmetry \cite{freidel-louapre-diff} (see also \cite{Noui:2004iy} for a similar mechanism for the physical inner product of $(2+1)$ gravity).

The AL projective limit construction \cite{al} leads to the notion of a configuration space $\overline{\cal A}$ as an (quantum) extension of the space of connections. Similarly we expect that the present construction will lead to an (quantum) extension $\overline{\cal E}$ of the space of fluxes. This has been attempted in \cite{flux3} with the help of the non--commutative flux representation \cite{flux}, however still using the standard LQG embedding. This work showed that in this case one rather encounters an inductive instead of a projective limit for the space of fluxes. This is intimately related to how holonomies and fluxes behave under the various refinement maps. The standard LQG embeddings lead to a composition of holonomies, whereas the condition on the fluxes are that these stay constant \cite{flux3}. The new BF embedding map composes fluxes, corresponding to the definition of integrated flux observables. On the other hand holonomies are refined so that curvature is left invariant.  Thus the BF embedding leads to a picture where fluxes are coarse grained geometrically and could hence be useful for developing geometric coarse graining and renormalization procedures related to the methods of \cite{group}.

 \section*{7. Conclusions} 

We sketched the construction of an alternative vacuum for loop quantum gravity, which is dual to the standard Ashtekar--Lewandowski vacuum. Since this BF vacuum incorporates also non--degenerate geometries, we expect that it will facilitate very much the extraction of large scale physics from loop quantum gravity, and develop further its applications to cosmology. Future work will make many notions more precise and provide also the construction for the $(3+1)$--dimensional case. In this case one will have to use 1--4 and 2--3 refining Pachner moves. The cylindrically consistent observables will be again closed holonomies and integrated fluxes transported to a root. These integrated fluxes will however describe the addition of normals belonging to pieces of a surface, thereby implementing a geometric composition of surfaces related to a 2--category structure \cite{baratin}.

Our construction requires the modification of certain ingredients of loop quantum gravity, in particular the introduction of triangulations\footnote{The inclusion of generalized triangulations such as quadrangulations seems  to be straightforward, since these can be obtained from the appropriate gluing of triangles.} as opposed to just one--complexes (graphs). This necessity was also discussed in \cite{matteo-valentin} in the context of a path integral quantization of BF theory. Although we expect important changes in the standard framework of quantum gravity with the adaptation of this new vacuum, many central results remain the same, notably the discreteness\footnote{Certain ordering choices for the $\SU(2)$ Casimir, that are disfavored by the AL cylindrical consistency, might now be allowed by BF cylindrical consistency.} of the spectra of geometric operators \cite{geometric-ops}. To study this more carefully, one needs the (integrated) fluxes themselves as operators. Choosing a (Bohr) compactification for the dual will however only allow to use the exponentiated fluxes as operators.

The new representation presented here is unitarily inequivalent to the AL representation. One might wonder how this new representation evades the conclusion of the F--LOST uniqueness theorem \cite{FLOST}, stating that the AL representation is the only representation of the holonomy flux algebra, satisfying a number of conditions. As mentioned one point is that the holonomy flux algebra is changed. In particular, it is essential to use ``simplicial'' fluxes, which also incorporate the parallel transport. A precise definition of the observable algebra underlying the new representation is presented in \cite{to appear}. Additionally the new representation only allows exponentiated fluxes, which will not lead to   weakly continuous families. In fact in the case of the $\text{U}(1)$ structure group we encounter the Bohr compactification of the dual ${\mathbb Z}$ of $\text{U}(1)$. This requires the exponentiation of the fluxes, as is known from loop quantum cosmology \cite{LQC} in the case of operators encoding the holonomy (and allows to evade the Stone- von Neumann uniqueness theorem for quantum mechanics). On a more heuristic level, the fact that the vacuum state is given by group delta functions forbids the action of fluxes as derivative operators and requires exponential fluxes that translate the arguments of the delta functions, as described above.

We can envisage many generalizations. This framework might facilitate for example the construction of a Hilbert space based on non--compact groups (which is required for Lorentzian signature in three dimensions and for the self-dual Lorentzian theory in four dimensions) \cite{eteraSL}. Indeed, the AL cylindrical consistency requires amenable groups, which is not the case for $\text{SL}(2,\mathbb{C})$. Furthermore, similar to the construction of Koslowski, which shifts the expectation value of the fluxes, we can try to obtain a vacuum peaked on homogeneous instead of flat connections. This opens applications to loop quantum cosmology \cite{LQC}, including that of a possibly time varying homogeneous connection, but also relates to the question of whether one can derive a quantum group structure for loop quantum gravity \cite{Qgroups}. Quantum groups with deformation parameter at the root of unity have in a certain sense a compact dual corresponding to an exponentiated flux observable (for the torus subgroup). 

Finally, let us comment on possible definitions of Hamiltonian dynamics. The refinement by Pachner moves does indeed suggest an implementation of the dynamics via Pachner moves, as discussed in \cite{hoehn,refining}. This would nicely connect to the covariant spin foam picture \cite{CSF}. Nevertheless, also an infinitesimal implementation of the dynamics via Hamiltonian constraints seems to be possible. The construction of Thiemann \cite{thiemann} based on the AL vacuum is said to lead to a finite result since the Hamiltonian acts only on (dual) nodes where volume is concentrated. A similar mechanism can be expected for the BF vacuum, where the Hamiltonian will only act on triangulation vertices (or edges in four dimensions), where curvature is concentrated (see also \cite{zipfel}). This addresses a criticism of Immirzi \cite{immirzi} towards the LQG Hamiltonian, and corresponds more directly to diffeomorphisms as vertex translations, as it is implemented without (discretization) anomalies (at least at the level of classical discretizations) for various flat and homogeneous geometries \cite{bonzomdittrich} (see \cite{bonzomfreidel} for the quantization).

The question of the constraint algebra will however remain open for cases with propagating degrees of freedom \cite{bahrdittrich}. Anomalies resulting from discretization artifacts can be avoided with an exact or perfect discretization, leading to a possibly non--local Hamiltonian \cite{dittrich12r,refining} (see also the discussion in \cite{ziprick}). Alternatively, a renormalization procedure might lead to a restoration of diffeomorphism symmetry in the continuum limit \cite{improved,refining}. In this respect, the work presented here realizes for $(2+1)$--dimensional gravity the construction of a continuum limit of loop quantum gravity (and spin foam quantization) via dynamical cylindrical consistency, which has been outlined in \cite{bd12b,refining}. Thus, we can hope that such a construction leading to a physical vacuum is applicable for $(3+1)$--dimensional gravity as well.

\section*{Acknowledgements}

We thank Valentin Bonzom, Wojciech Kami{\'n}ski and Lee Smolin for discussions, and Laurent Freidel for providing the draft \cite{ribbons}. MG thanks Perimeter Institute for hospitality during the final stages of this work. This research was supported by Perimeter Institute for Theoretical Physics. Research at Perimeter Institute is supported by the Government of Canada through Industry Canada and by the Province of Ontario through the Ministry of Research and Innovation. MG is supported by the NSF Grant PHY-1205388 and the Eberly research funds of The Pennsylvania State University.

\end{document}